\documentclass[sigconf,natbib=true]{acmart}

\setcopyright{acmlicensed}
\copyrightyear{2025}
\acmYear{2025}
\acmDOI{XXXXXXX.XXXXXXX}
\acmConference[Conference acronym 'XX]{Conference title}{June 03--05, 2025}{Woodstock, NY}
\acmISBN{978-1-4503-XXXX-X/2025/06}

\usepackage{balance}
\usepackage{amsfonts}
\usepackage{multirow}
\usepackage{pifont}
\usepackage{makecell}
\usepackage{subcaption}
\usepackage{colortbl}
\usepackage{xcolor}
\usepackage{tabularx}
\usepackage{soul}
\usepackage{algorithm}
\usepackage[noend]{algorithmic}
\usepackage{booktabs}
\usepackage{bm}
\usepackage{amsmath}
\definecolor{darkred}{rgb}{0.875, 0, 0}
\definecolor{darkgreen}{rgb}{0, 0.5, 0}

\begin{document}
\title{Comparative Explanations via Counterfactual Reasoning in Recommendations}
\author{Yi Yu}
\email{yiyu52@huawei.com}
\affiliation{
  \institution{Huawei Technologies Co., Ltd.}
  \city{Beijing}
  \country{China}
}

\author{Zhenxing Hu}
\email{huzhenxing1@huawei.com}
\affiliation{
  \institution{Huawei Technologies Co., Ltd.}
  \city{Beijing}
  \country{China}
}

\begin{abstract}
Explainable recommendation through counterfactual reasoning seeks to identify the influential aspects of items in recommendations, which can then be used as explanations. However, state-of-the-art approaches, which aim to minimize changes in product aspects while reversing their recommended decisions according to an aggregated decision boundary score, often lead to factual inaccuracies in explanations. To solve this problem, in this work we propose a novel method of \textbf{Co}mparative \textbf{Count}erfactual \textbf{E}xplanations for \textbf{R}ecommendation (CoCountER). 
CoCountER creates counterfactual data based on soft swap operations, enabling explanations for recommendations of arbitrary pairs of comparative items. Empirical experiments validate the effectiveness of our approach.
\end{abstract}


\maketitle

\section{Introduction}
Recommender systems that offer both high-quality recommendations and clear explanations tend to be more appealing to users, as these explanations can help users understand the underlying rationale behind recommendations, leading to greater trust and satisfaction among users with the system~\cite{tintarev2007survey, chen2019user, zhang2020SvyExplainable, ooge2022explaining, lu2023user, yu2024sequential, yu2025beyond}. 

\begin{figure}[t]
    \centering
    \includegraphics[width=0.48 \textwidth, height=0.175\textheight]{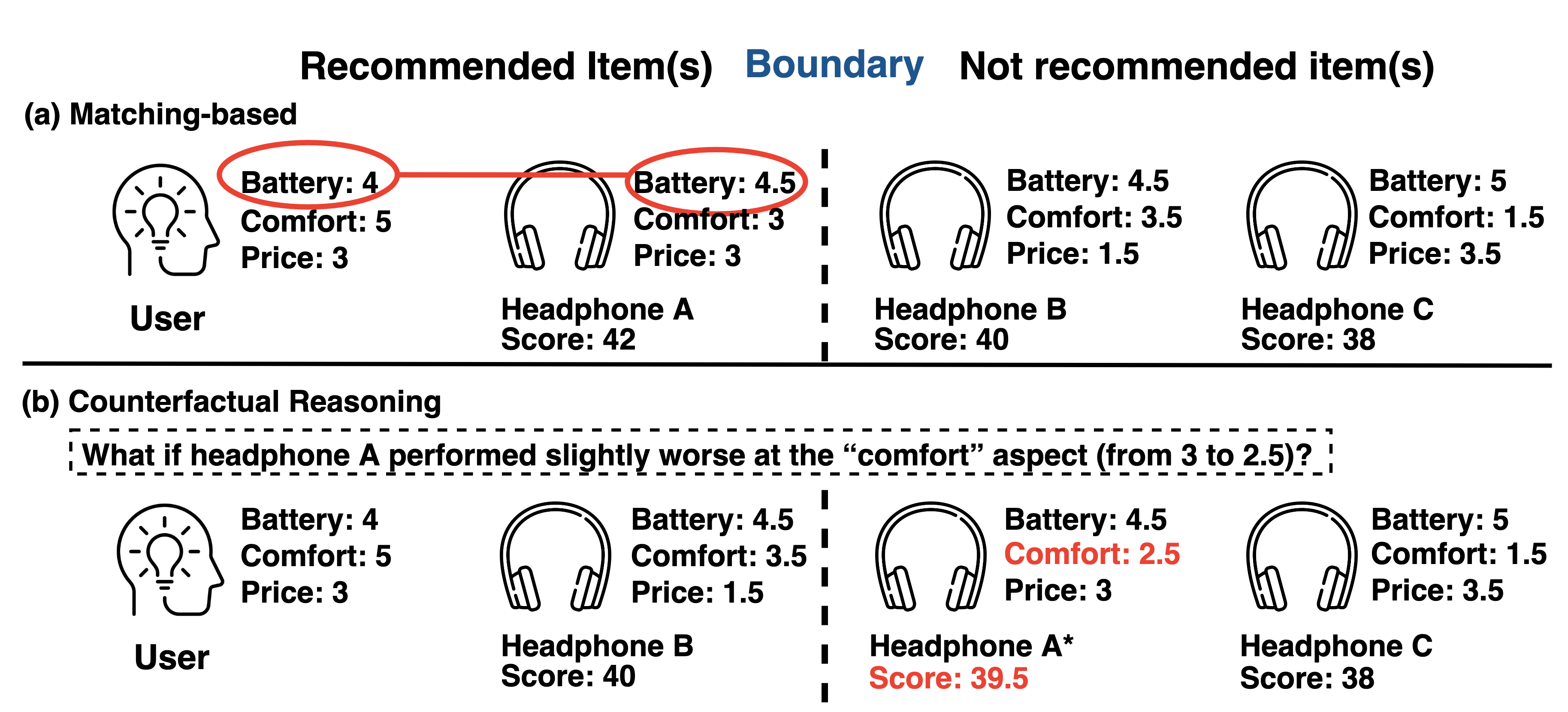}
    \caption{Matching-based versus counterfactual reasoning methods. The numerical values of the aspects reflect the user's focus on each aspect and the item's performance in each aspect, which are derived from user-reviews.}
    \label{fig:figure1}
\end{figure}

One primary approach utilizes explicit item aspects (attributes) to construct template-like explanations: “You might be interested in [feature], in which this product excels.” Models like EFM~\cite{zhang2014EFM}, MTER~\cite{wang2018MTER}, and A2CF~\cite{chen2020try} adopt tensor-factorization techniques to simultaneously optimize recommendations and aspect representations for users/items,  which are also known as matching-based methods. In contrast, the application of causal reasoning in machine learning sheds light on explainable recommendation from counterfactual perspective. Tan et al.~\cite{tan2021counterfactual} proposed CountER based on a counterfactual reasoning framework with the intuition of minimally reducing input features to reverse the decisions, thus identifying the signals that trigger recommendations. We use an example illustrated in Figure~\ref{fig:figure1} to discuss the differences in explanations by these two methods. 

From the figure, each user has assigned an attention score to every aspect, while each item receives a performance score for each aspect, similar to~\cite{tan2021counterfactual}. The recommender system ranks items based on their total score. For instance, the total score of Headphone A is $4\times4.5+5\times3+3\times3 = 42$, positioning it at the top of the recommendation list. To explain the recommendation of Headphone A, a matching-based model would attribute it to the \textit{battery} aspect due to its highest multiplication score of 18 ($4\times4.5$), 
surpassing \textit{comfort} ($5\times3=15$) and \textit{price} ($3\times3=9$). On the other side, a counterfactual reasoning approach would identify \textit{comfort} as the explanation, as it requires only a slight reduction from 3 to 2.5 to reverse the recommendation decision, whereas changing \textit{battery} or \textit{price} necessitates more significant alterations. Hence, \textit{comfort} serves as the primary triggering factor for the recommendation.
\textbf{Notably, we can observe factual inaccuracies in the explanations provided by these methods:} (i) in the matching-based model, \textit{battery} is erroneously identified as the explanation despite it actually performs
the worst on \textit{battery} among all products. This indicates that aspects with high matching scores may not consistently be the rationale behind the recommendation. (ii) in the counterfactual reasoning model, choosing \textit{comfort} as an explanation is acceptable when comparing Headphone A with C. However, when comparing it with Headphone B, we find that Headphone A is indeed inferior in \textit{comfort}. This implies that existing counterfactual methods also result in explanation errors that contradict intuition. Through theoretical analysis, we identify the issue lies in the model optimization process, which aims to push the total sum of reduced aspects to the decision boundary score (i.e., score=40 by Headphone B), rather than pushing each aspect to the decision boundary.

\begin{figure}[t]
    \centering
    \includegraphics[width=0.48 \textwidth, height=0.178\textheight]{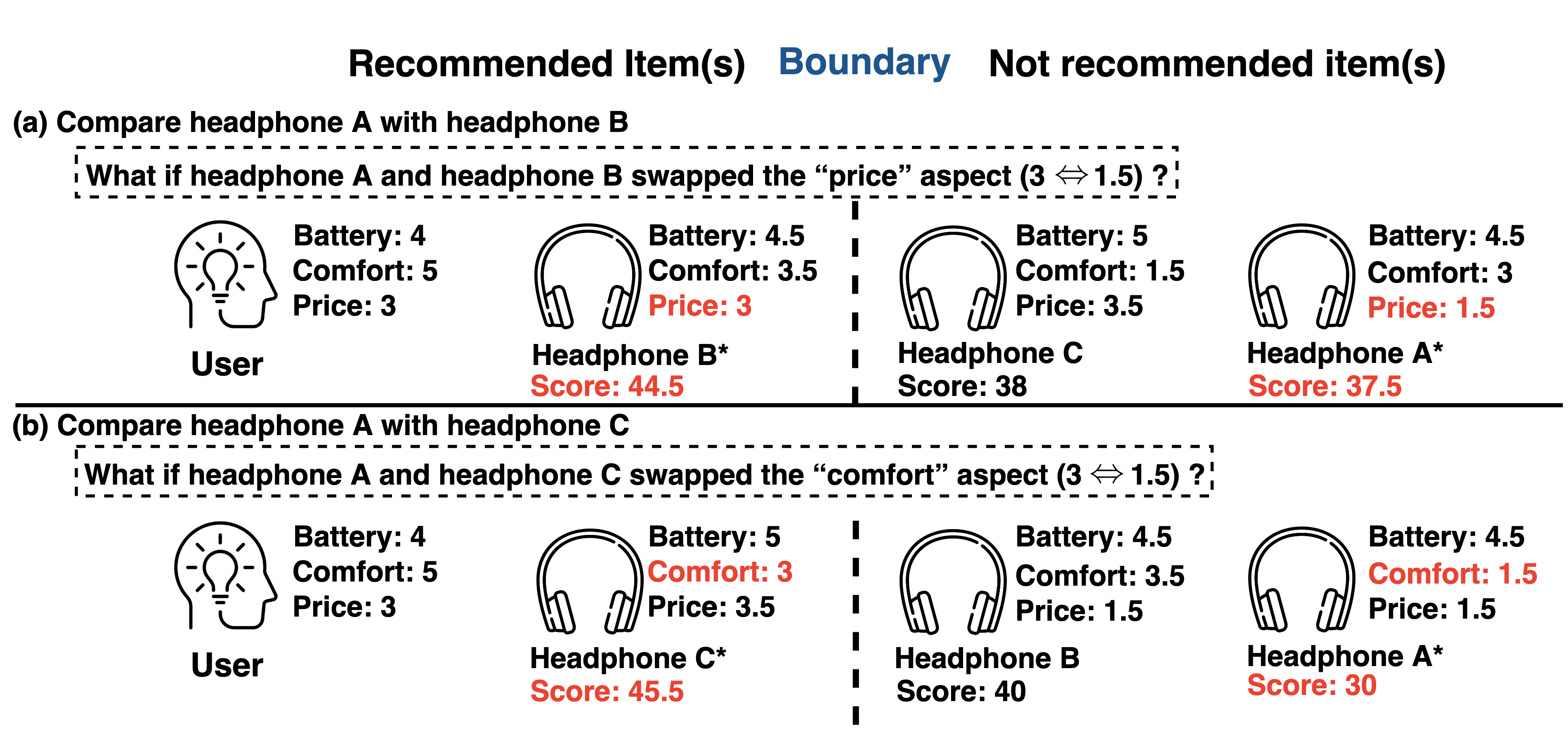}
    \caption{Our proposal of comparative counterfactual reasoning method. In this example, to explain the Headphone A, first select the reference item (B or C), then intervene by swapping the aspect values between comparative items to alter their relative rankings.}
    \label{fig:figure2}
\end{figure}

We tackle the aforementioned issue by employing counterfactual reasoning to compare aspects between pairs of items. To achieve this, we propose a novel counterfactual operations, \textbf{swap operation}. In contrast to previous reduction-based method, swap operation allows manipulation at the aspect level for a pair of comparative items, which consist of an explaining item and a reference item.
Using the same example, we illustrate the intuition of our method in Figure~\ref{fig:figure2}. To explain the Headphone A, we first select the reference item, such as Headphone B as depicted in the upper part of the figure. Then, we optimize for the minimal swap operations required on their respective aspects to ensure that Headphone A is positioned after B in the recommendation list. Hence, \textit{price} is chosen as the explanatory factor because swapping their \textit{price} values can alter the output (while \textit{battery} and \textit{comfort} cannot). This demonstrates that \textit{price} is a highly influential factor in recommending Headphone A over B. 
Likewise, in the comparison with Headphone C, we can see that \textit{comfort} is the key factor for recommending Headphone A over C. Therefore, it is evident that the influential aspects for recommendations generated through the comparative manner are more faithful.

We make the following contributions in this work: (1) We propose a novel approach for creating counterfactual data through our swap operation. (2) We propose comparative counterfactual explanations, enabling explaining recommendations for given arbitrary pair of comparative items. (3) We experimentally demonstrate the effectiveness of our approach.

\vspace{-0.5em}
\section{Related Work}
Existing approaches on explainable recommendation systems utilize review texts to develop aspect-aware interpretable representations of users and items. While many works~\cite{mcauley2013HFT, zhang2014EFM, wang2018MTER, chen2020try} employ a multi-task joint tensor factorization framework to derive features for users and items, Tan et al.~\cite{tan2021counterfactual} argue that these algorithms essentially align with a matching framework. They for the first time generate explanations from a counterfactual perspective: “what would occur if we intervene these signals
to alternative values”, thus establishing a counterfactual reasoning framework, \textbf{CountER}. Later, Ranjbar et al.~\cite{ranjbar2023explaining} integrates textual features extracted from a pre-trained large language model such as BERT~\cite{kenton2019bert} into the counterfactual framework. Yu et al.~\cite{yu2023adarex, yu2025domain} incorporate a causal inference perspective to generate counterfactual examples across different domains, a method referred to as domain counterfactuals.
Notably, the counterfactual samples not only serves the purpose of explanation but also is utilized for data augmentation to enhance the predictive accuracy of the recommendation model, as demonstrated in~\cite{xiong2021counterfactual, yu2023counterfactual, wang2023improving, boratto2023counterfactual, ji2023counterfactual}. Inspired by the comparative explanation proposed by Yang et al.~\cite{yang2022comparative}, we argue that elucidating how one item is compared with another is more persuasive and rational than isolatedly providing explanations. Yet, our work differs significantly from that of \cite{yang2022comparative}: we provide explanations by intervening on features through the counterfactual reasoning viewpoint, in line with CountER framework~\cite{tan2021counterfactual}. In contrast, Yang et al. \cite{yang2022comparative} employs an autoregressive decoder on user-item hidden states to generate explanation texts, adhering to the NRT framework~\cite{li2017neural}. In summary, we propose the pioneering comparative counterfactual reasoning method for explaining recommendations, unveiling a novel category of counterfactuals.

\vspace{-0.5em}
\section{Methodology}
\subsection{Preliminaries}
Let $\mathcal{U}$, $\mathcal{I}$, and $\mathcal{R}$ represent the sets of users, items, and reviews, respectively. Consistent with \cite{xiong2021counterfactual, tan2021counterfactual, chen2020try}, we utilize a well-established sentiment analysis toolkit, Sentires\footnote{\url{https://github.com/evison/Sentires}}~\cite{zhang2014Sentires} to extract aspect-sentiment pairs. It yields a sentiment lexicon for the entire dataset, with each entry $(user, item, attribute, sentiment)$, denoted as $(u, i, a, s)$. The sentiment $s$ is either $+1$ or $-1$, signifying the user's satisfaction regarding a specific aspect, and the collection of all users concerning that aspect reveals the item's performance in that aspect.
Using the entries extracted from the dataset, we construct the user-aspect attention matrix $X$, and the item-aspect quality matrix $Y$. Here, $X_{u, a}$ denotes user $u$'s attention in aspect $a$, while $Y_{i, a}$ reflects item $i$'s quality in aspect $a$, calculated as follows:
{\small
\begin{equation}
\begin{aligned}
& X_{u, a}=\left\{\begin{array}{l}
0, \text { if user } u \text { has not mentioned aspect } a \\
1+(N-1)\cdot \frac{1-\exp^{-t_{u,a}}}{1+\exp^{-t_{u,a}}}
, 
\,\, \text {otherwise}
\end{array}\right. \\
& Y_{i, a}=\left\{\begin{array}{l}
0, \text { if item } i \text { has not mentioned on aspect } a \\
1+\frac{N-1}{1+\exp^{-t_{i, a} \cdot \bar{s}_{i, a}}}, \,\, \text {otherwise}
\end{array}\right.
\end{aligned}
\end{equation}
}
where $t_{u, a}$ is the total count of user $u$ mentioning aspect $a$. $N$ is the scaling factor such that $1 \leq  X_{u, a} \leq N$, which is often set as the highest rating score within the recommender platform~\cite{zhang2014Sentires, xiong2021counterfactual, tan2021counterfactual, chen2020try}.  $t_{i,a}$ is the frequency that item $i$
is mentioned on aspect $a$, while $\bar{s}_{i,a}$
is the average of all the sentiment scores in those $t_{i,a}$ mentions.

Next, we train a recommendation model $g_{\theta}(\cdot)$ to serve as the black-box model we aim to explain, where $g_{\theta}(\cdot)$ predicts the user-item affinity score based on user and item representations $X_u, Y_i$:
{\small
\begin{equation}
    r_{u,i} = g_{\theta}(X_u, Y_i), 
\end{equation}}
where $X_u, Y_i \in \mathbb{R}^{k}$ and $k$ is the size for aspect set. 

Since our main focus is not on the recommendation accuracy but on the explanability, we simply employ a fusion layer to concatenate the user's and the item's features and then pass through three fully connected layers with ReLU as activation function and output through a Sigmoid function, as suggested by \cite{tan2021counterfactual}. We train the recommendation model using a cross-entropy loss function, with a 1:2 ratio of positive (label 1, user interacted with the item) to negative (label 0) samples. The recommendation model $g_{\theta}(\cdot)$ can be more sophisticated, for instance, adopting a Graph Neural Network (GNN) -based model, provided it integrates aspect features of both users and items into its inputs.

\subsection{Comparative Counterfactual Explanations}
For a target user $u$, considering a comparative pair of items including an explaining item $i$ and a reference item 
$j$, their recommendation scores $r_{u,i}$ and $r_{u,j}$
should satisfy the following requirement (the selection criteria for reference item $j$ is discussed in Section~\ref{sec:discussion}):
{\small 
\begin{equation}
    r_{u,i} > r_{u,j},
\end{equation}
}

\noindent where $r_{u,i}=g_{\theta}(X_u, Y_i), r_{u,j}=g_{\theta}(X_u, Y_j)$.
Our method identifies the most influential features affecting the recommendation decision by performing minimal interventions—specifically, swap operations on pairwise aspect values between comparative items, to reverse their relative rankings. Mathematically, we formulate it as an optimization problem as follows:
{\small
\begin{equation}
\begin{aligned}
 \text{Minimize}& \sum \text{swap variables } \boldsymbol{\psi} \\
s.t., &r_{u,i^*} < r_{u,j^*}, \\
&r_{u,i^*} = g_{\theta}(X_u, f(Y_i, Y_j, \boldsymbol{\psi})), \\
&r_{u,j^*} = g_{\theta}(X_u, f(Y_j, Y_i, \boldsymbol{\psi})),
\end{aligned}
\end{equation}
}

\noindent where $f(\cdot)$ denotes the swap function, $r_{u,i^*}, r_{u,j^*}$ are the ranking scores for synthesized counterfactual items $i^*$ and $j^*$ after swapping, respectively. 

A significant challenge here is defining swap functions $f(\cdot)$ in a manner that allows for differentiability, thereby enabling the problem to be solved using automatic differentiation tools through stochastic gradient descent~\cite{ruder2016overview}.
Intuitively, swapping two variables can be seen as performing the following calculations on both variables simultaneously: multiplying itself by 0 and then adding the other multiplied by 1. To achieve a soft control of swap operations, we utilize a sigmoid function $\sigma$, which ensures that the swap operation's value spans from 0 to 1. This trick offers an intuitive interpretation, where a value close to 0 indicates no swap, and a value approaching 1 indicates a complete swap. Therefore, we can define the swap function $f(\cdot)$ between comparative items $i$ and $j$ for generating counterfactual item features as follows:
{\small
\begin{equation}~\label{eq:swap}
    f(Y_i, Y_j, \boldsymbol{\psi}) = 
\begin{pmatrix}
1-\sigma(\boldsymbol{\psi}_1) \\
1-\sigma(\boldsymbol{\psi}_2) \\
\vdots \\
1-\sigma(\boldsymbol{\psi}_k) 
\end{pmatrix}
\cdot
\begin{pmatrix}
Y_{i1} \\
Y_{i2}\\
\vdots \\
Y_{ik}
\end{pmatrix}+
\begin{pmatrix}
\sigma(\boldsymbol{\psi}_1) \\
\sigma(\boldsymbol{\psi}_2) \\
\vdots \\
\sigma(\boldsymbol{\psi}_k) 
\end{pmatrix}
\cdot\begin{pmatrix}
Y_{j1} \\
Y_{j2} \\
\vdots \\
Y_{jk},
\end{pmatrix}
\end{equation}
}


Next, we can safely express the constraint of reversed relative ranking between comparative items as a margin ranking loss:
{\small
\begin{equation}
     \mathcal{L}(r_{u,i^*}, r_{u,j^*}) = \max(0, (r_{u,i^*} - r_{u,j^*})+m),
\end{equation}
}

where $m$ is the margin threshold for error tolerance in ranking. Incorporating this constraint as a Lagrange multiplier~\cite{zhang2014EFM} leads to our final optimization objective:
{\small
\begin{equation}~\label{eq:goal}
    \min_{\boldsymbol{\psi}} ||\sigma(\boldsymbol{\psi})||_1 + \lambda \mathcal{L}(r_{u,i^*}, r_{u,j^*}),
\end{equation}}

where $\lambda$ is the scaling factor to balance the two objectives.
Optimizing this formula yields $\boldsymbol{\psi}^{'}$, representing the optimal swap variables. The explanations of aspects are identified as the ones whose swap operations successfully reverse the comparative items' relative ranking, with a threshold set at 0.5 for identifying swaps:
{\small
\begin{equation}~\label{eq:aspects}
     \hat{\mathcal{E}} = \{a|\sigma(\boldsymbol{\psi}{_a}) > 0.5\}.
\end{equation}
}

\subsection{Model Optimization}
We analyze how CoCountER updates the parameters $\boldsymbol{\psi}$. With the optimization objective in Eq.~\ref{eq:goal} denoted as $O(\mathcal{\boldsymbol{\psi}})$, the update rule for $\boldsymbol{\psi}$ in gradient descent is:  
{\small
\begin{equation}~\label{eq:update1}
    \boldsymbol{\psi}^{(new)} = \boldsymbol{\psi}^{(old)} - \eta\nabla_{\boldsymbol{\psi}} O(\boldsymbol{\psi})
\end{equation}}

\noindent where $\eta$ is the learning rate, and $\nabla_{\boldsymbol{\psi}} O(\boldsymbol{\psi})$ is the gradient of the optimization objective with respect to $\eta$. For every component $\boldsymbol{\psi}_a$ within $\boldsymbol{\psi}$, its partial derivative, which is the sum of two components, is computed as follows:
{\small
\begin{equation}~\label{eq:update2}
\begin{aligned}
\nabla_{\boldsymbol{\psi}_a} O(\boldsymbol{\psi}_a) &= \frac{\partial}{\partial \boldsymbol{\psi}_a}||\sigma(\boldsymbol{\psi}_a)||_1 + \frac{\partial}{\partial \boldsymbol{\psi}_a} \lambda \mathcal{L}(r_{u,i^*}, r_{u,j^*}) \\
&= \frac{\partial}{\partial \boldsymbol{\psi}_a} \sum_{a=1}^{k}\sigma(\boldsymbol{\psi}_a) + \frac{\partial}{\partial \boldsymbol{\psi}_a} \lambda (r_{u,i^*}-r_{u,j^*}) \\
&= \sigma(\boldsymbol{\psi}_a)(1-\sigma(\boldsymbol{\psi}_a)) \\
&\quad + \lambda \frac{\partial}{\partial \boldsymbol{\psi}_a} g_{\theta}(X_u, Y_i-Y_i\sigma(\boldsymbol{\psi})+Y_j\sigma(\boldsymbol{\psi})) \\
&\quad - \lambda \frac{\partial}{\partial \boldsymbol{\psi}_a}g_{\theta}(X_u, Y_j-Y_j\sigma(\boldsymbol{\psi})+Y_i\sigma(\boldsymbol{\psi}))
\end{aligned}
\end{equation}}

Given that $g_{\theta}(\cdot)$ is assumed to be a black-box recommendation model, one can incorporate the results from Eequation~(\ref{eq:update2}) into it to obtain the corresponding model's final gradient; thus, we omit further derivation here. Alternatively, since $g_{\theta}(\cdot)$ is generally implemented using neural networks, one can leverage automatic differentiation tools such as PyTorch~\footnote{\url{https://pytorch.org/}} for this purpose.
We provided the full process of our methodology to generate explanations for recommendations in Algorithm~\ref{alg:the_alg}.
\begin{algorithm}[t]
\caption{CoCountER: Comparative Counterfactual Explanations for Recommendation}
\label{alg:the_alg}
\begin{algorithmic}[1]
\small 
\STATE\textbf{Input:} Trained rec model $g_{\theta}(\cdot)$, user $u$, recommended item $i$.
\STATE\textbf{Output:} Explanation set \( \mathcal{E} \) justifying item $i$ for user $u$.
\STATE Initialize the explanation set $\mathcal{E} = \emptyset $.
\STATE Determine the reference item list $J =\{j_1, j_2, ..., |r_{u,j} < r_{u,i}\}$.
\FOR{$j$ in $J$}
    \STATE Randomly initialize the trainable vector $\boldsymbol{\psi}$
    \STATE Optimize for Eq.~(\ref{eq:goal}) using Eq.~(\ref{eq:update1}, \ref{eq:update2})
    \IF{$r_{u, j^*} > r_{u,i^*}$}
        \STATE Obtain $\hat{\mathcal{E}}$ by Eq.~(\ref{eq:aspects}).
    \ENDIF
    \FOR{$e \in \hat{\mathcal{E}}$}
        \IF{$e \not\in \mathcal{E}$}
            \STATE $\mathcal{E} \gets \mathcal{E} \cup \{e\}$
        \ENDIF
    \ENDFOR
\ENDFOR
\end{algorithmic}
\end{algorithm}

\subsection{Discussion}~\label{sec:discussion}
\noindent\textbf{Relationship with Reduction-based Counterfactual Methods.} Our counterfactual method utilizes swap operations. It degenerates into reduction-based when the following actions are taken: (i) using a fixed reference item, and (ii) only computing the first part of Equation~(\ref{eq:swap}) (i.e., $(1-\sigma(\boldsymbol{\psi}))\cdot Y$). Therefore, our method represents a more generalized form. This suggests that future research may explore incorporating various operations into counterfactual frameworks to enhance their versatility and effectiveness.

\noindent{\textbf{Position and Number of Reference Items.}}
Two factors affect the aspect generation for a recommended item.
(i) \textit{Position of the reference item:} as shown in Equation~(\ref{eq:goal}), the optimization becomes simpler when the reference item $j$ is ranked lower (i.e., has a smaller recommendation score $r_{u,j}$), allowing more aspects to satisfy the criteria. Consequently, lower-ranked reference items yield more potential explanations. For example, comparing the top-ranked item with one ranked 20th generally provides clearer rationales than comparing it with the 2nd-ranked item.
(ii) \textit{Number of reference items:} since different references produce distinct aspects, incorporating more reference items increases computational cost but improves the recall of explanations. These factors are further analyzed in the experiments.
\section{Experiments}
\subsection{Experimental Setup}
\noindent{\textbf{Data Preparation.}}
\begin{table}[t]
\centering
 \caption{Statistics of the datasets. “\# exp./review” denotes the average number of explanations per review.}
 \renewcommand{\arraystretch}{0.95} 
 \resizebox{0.47 \textwidth}{!}{%
 \setlength{\tabcolsep}{2.5pt}
\begin{tabular}{l l l l l l l}
\specialrule{1pt}{0pt}{1pt}
    \textbf{Dataset} & \textbf{\# users} & \textbf{\# items} & \textbf{\# reviews} & \textbf{\# aspects} & \textbf{\# exp./review} & \textbf{density (\%)} \\\hline
    Electronics & 963 & 1,112 & 19,418 & 877 & 2.70 & 1.813 \\
    CDs \& Vinyl & 2,129 & 2,907 & 56,045 & 810 & 1.89 & 0.906 \\
    Movies & 5,586 & 6,703 & 187,490 & 1,530 & 1.80 & 0.501 \\
\specialrule{1pt}{0pt}{1pt}
  \end{tabular}}
 \label{tbl:statistics}
\end{table}
We conduct experiments using Amazon review datasets~\cite{he2016amazon_review}.\footnote{\url{https://jmcauley.ucsd.edu/data/amazon/}} We utilize three categories: \textit{Office}, \textit{CDs and Vinyl}, and \textit{Movies}. To preprocess the data, following~\cite{tan2021counterfactual}, we iteratively filter out users and items with fewer than 10 interactions and those interactions with empty explanations. Each dataset is randomly split into training, validation, and test sets with an 8:1:1 ratio. Dataset statistics are presened in Table~\ref{tbl:statistics}.

\noindent{\textbf{Compared Algorithms.}}
We compare the following four types of methods: (i) \textit{Random}: We randomly select aspects from the aspect space as explanations, which serves as a indicator of task difficulty. (ii) \textit{Sort}: For each product recommended to the user, we sort the item performance on aspects and select the highest performing aspects as explanations, referred to as \textbf{Sort-i}. Similarly, we sort based on user attention scores on aspects, denoted as \textbf{Sort-u}. (iii) \textit{Matching-based}: \textbf{EFM}~\cite{zhang2014EFM} and \textbf{A2CF}~\cite{chen2020try}.
EFM combines explicit aspect features and implicit features to predict recommendations, and A2CF captures user-item-feature correlations using an attentive neural network. For both models, we select the aspects with the highest matching scores as explanations for recommended items. 
(iv) \textit{Counterfactual-based}: \textbf{CountER}~\cite{tan2021counterfactual} and our \textbf{CoCountER}. CountER is the first framework to employ counterfactual reasoning in explainable recommendation, and our CoCountER utilizes a comparative approach to counterfactual reasoning, allowing for comparisons between different items to generate explanations.

\noindent{\textbf{Evaluation Metrics.}}
For evaluating the explainability, we adopt \textit{user-oriented} measures, Recall and Precision, and \textit{model-oriented} evaluation measures, Probability of Necessity (PN) and Probability of Sufficiency (PS)~\cite{tan2021counterfactual, pearl2009causal, ji2023counterfactual}.
Precision measures the percentage of predicted explanations that match the ground truth, while Recall measures the percentage of ground truth explanations that are included in the predictions. The PN metric evaluates the necessity of specific aspects for decision-making by removing them and verifying if recommendations remain effective, while the PS metric calculates whether these aspects alone are sufficient for decision-making by exclusively using them for recommendations. In all of these metrics, higher values indicate better performance.
\subsection{Results and Analysis}
\subsubsection{Experimental Results}

\begin{table}[t]
\centering
\caption{Overall performance comparison of the evaluated methods. 
PR, RC, PN, and PS denote Precision, Recall, Probability of Necessity, 
and Probability of Sufficiency, respectively. 
Bold and underlined values indicate the best and second-best results, respectively.}
\label{tbl:results}

\renewcommand{\arraystretch}{0.92}
\setlength{\tabcolsep}{10pt}
\footnotesize

\begin{tabularx}{0.42\textwidth}{l *{4}{>{\centering\arraybackslash}X}}
\specialrule{1pt}{0pt}{1pt}
\multicolumn{5}{c}{\textbf{Electronics}} \\
\hline
\textbf{Methods} & \textbf{PR} & \textbf{RC} & \textbf{PN} & \textbf{PS} \\
\hline
Random & 0.001 & 0.001 & 0.010 & 0.075 \\
Sort-i & 0.031 & 0.036 & 0.141 & 0.698 \\
Sort-u & 0.059 & 0.068 & 0.097 & 0.783 \\
EFM & \underline{0.086} & 0.150 & 0.325 & 0.888 \\
A3CF & 0.076 & 0.142 & 0.302 & \underline{0.908} \\
CountER & 0.081 & \textbf{0.168} & \underline{0.511} & 0.894 \\
CoCountER & \textbf{0.089} & \underline{0.157} & \textbf{0.734} & \textbf{0.931} \\
\hline
\multicolumn{5}{c}{\textbf{CDs \& Vinyl}} \\
\hline
\textbf{Methods} & \textbf{PR} & \textbf{RC} & \textbf{PN} & \textbf{PS} \\
\hline
Random & 0.002 & 0.002 & 0.010 & 0.075 \\
Sort-i & 0.038 & 0.043 & 0.158 & 0.719 \\
Sort-u & 0.057 & 0.066 & 0.102 & 0.794 \\
EFM & \textbf{0.096} & \underline{0.158} & 0.398 & 0.918 \\
A3CF & 0.082 & 0.121 & 0.472 & 0.912 \\
CountER & \underline{0.091} & \textbf{0.176} & \underline{0.526} & \underline{0.921} \\
CoCountER & 0.090 & 0.123 & \textbf{0.773} & \textbf{0.936} \\
\hline
\multicolumn{5}{c}{\textbf{Movies}} \\
\hline
\textbf{Methods} & \textbf{PR} & \textbf{RC} & \textbf{PN} & \textbf{PS} \\
\hline
Random & 0.001 & 0.001 & 0.010 & 0.075 \\
Sort-i & 0.029 & 0.034 & 0.118 & 0.681 \\
Sort-u & 0.052 & 0.071 & 0.089 & 0.786 \\
EFM & 0.083 & 0.145 & 0.312 & 0.882 \\
A3CF & \underline{0.087} & 0.146 & 0.297 & \underline{0.905} \\
CountER & 0.084 & \textbf{0.165} & \underline{0.496} & 0.889 \\
CoCountER & \textbf{0.090} & \underline{0.158} & \textbf{0.744} & \textbf{0.928} \\
\specialrule{1pt}{0pt}{1pt}
\end{tabularx}
\end{table}
The results in Table~\ref{tbl:results} show that CoCountER consistently surpasses all baselines across the three datasets in the counterfactual-related metrics, i.e., PN and PS. Compared with CountER, CoCountER achieves higher PN and PS, indicating a stronger ability to capture truly causal aspects behind recommendations. This improvement stems from our comparative counterfactual design: by performing aspect-level \textit{swap operations} between item pairs, CoCountER provides more faithful and context-aware explanations than the reduction-based strategy used in CountER.

\subsubsection{Hyper-parameter Analysis}
\begin{figure}[t]
    \centering
    \includegraphics[width=0.99\linewidth]{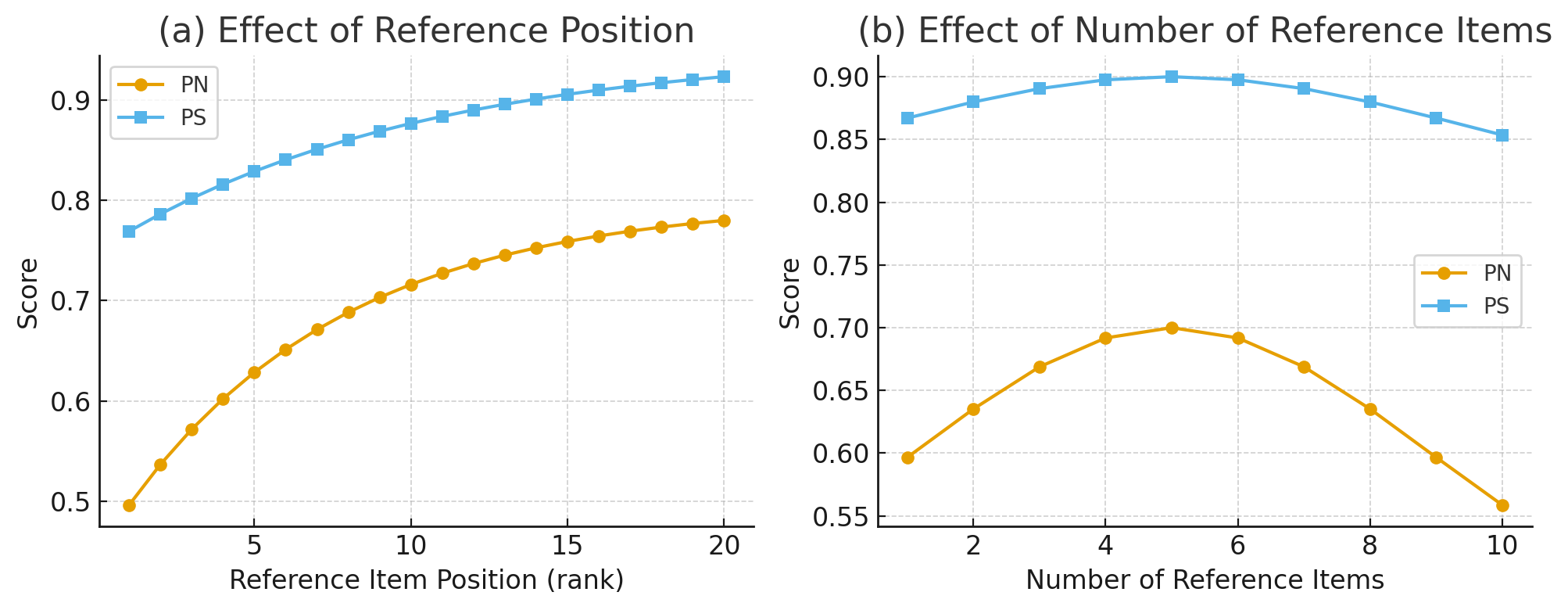}
    \caption{Hyper-parameter sensitivity analysis of CoCountER. 
    (a) Effect of reference item position. (b) Effect of the number of reference items. 
    Both PN and PS benefit from lower-ranked and moderately sized reference sets.}
    \label{fig:hyper}
\end{figure}

We analyze the impact of two key hyper-parameters: the \textit{position} of the reference item and the \textit{number} of reference items used in CoCountER. 
As shown in Figure~\ref{fig:hyper}, both factors noticeably affect the counterfactual metrics PN and PS. 
When the reference item appears lower in the ranking list, the optimization becomes less constrained, enabling the model to discover more valid counterfactual aspects—thus improving PN and PS (Figure~\ref{fig:hyper}(a)). 
Meanwhile, increasing the number of reference items initially enhances performance by introducing more diverse comparison signals, but excessive references lead to minor declines due to added noise (Figure~\ref{fig:hyper}(b)). 
Overall, CoCountER remains stable across a wide range of settings, demonstrating its robustness.

\section{Conclusion}
In this work, we proposed CoCountER, a comparative counterfactual framework that provides more faithful and causal explanations for recommendations.
Future work will explore integrating counterfactual reasoning with generative models to produce natural language explanations of counterfactual scenarios~\cite{li2021PETER, cheng-etal-2023-explainable, geng2022recommendation}.

\bibliographystyle{ACM-Reference-Format}
\balance
\bibliography{mybibliography}

\end{document}